# "Chaotic" kinetics, macroscopic fluctuations and long-term stability of the catalytic systems

Maria K. Koleva* and L. A. Petrov
Insititute of Catalysis, Bulgarian Academy of Science
1113 Sofia Bulgaria
Fax: +359 2 9712967
*e-mail: mkoleva@bas.bg

**Abstract**

Our recent interest is focused on establishing the necessary and sufficient conditions that guarantee a long-term stable evolution of both natural and artificial systems. Two necessary conditions, called global and local boundedness, are that a system stays stable if and only if the amount and the rate of exchange of energy and/or matter currently involved in any transition do not exceed the thresholds of stability of the system. The relationship between the local and global boundedness and the stability of the system introduces two new general properties of the state space and the motion in it, namely: the state space is always bounded, the successive steps of motion are always finite and involve only nearest neighbors. An immediate consequence of the boundedness is that the invariant measure of the state space is the normal distribution. The necessary condition for the asymptotic stability of the invariant measure is derived. It is found out that the state space exhibits strong chaotic properties regardless to the particularities of the system considered.

An example of kinetics that is compatible with both global and local boundedness is considered.

**Introduction**

A widespread conjecture of the statistical mechanics is that for the systems not far from the equilibrium there always exists a single state at which the system arrives asymptotically starting at any initial state. After that the system can undergo fluctuations so that the motion in its state space has an invariant measure. So far the only way to construct such general frame is based on the assumption that the state space of the system is connected by a finite irreducible Markov chain that always converges to a single stationary distribution over configurations. The motion on a phase trajectory involves transitions between any two states, far distant included (van Kampen, 1981).

Our goal now is to study how this conjecture is modified under considering the development of the fluctuations in the course of time.

The involvement of the time in the state space motion is immediately interrelated with the issue of the stability of the system, namely: the system stays stable if and only if the amount and the rate of exchange of energy and/or matter currently involved in any transition does not exceed the thresholds of stability of the system. To elucidate this limitation let us consider an example that comes from the study of anomalous fluctuations developed by the dynamics of the local elementary processes. The study of this type processes involves numerical "shooting" of the sequence of spatial configurations that develops in the course of the time under the dynamics induced by given local rules and geometrical constraints. Under the supposition that the state space is connected by a Markovian chain, the aim is to find out the stationary distribution to which the state space converges. Yet, our question now is whether the transitions between any two states are admissible, i.e. which transitions do not violate the stability of the systems?! Next we present an example that certain transitions can cause system breakdown. Among the great variety of anomalous fluctuations we consider the bistability first established by (Fitchtorn et al, 1989). Further it has been found out that it occurs both for finite (Liu and Evans, 1999) and infinite systems (James et al, 1999). In the latter case a stable reactive state (with high oxygen and low $N$ coverage) coexists with a stable inactive near $CO$-poisoned state ( high $CO$ and low oxygen coverage). Under the supposition that only short-range interactions are involved in the elementary processes of adsorption and reaction, the global transition goes trough local transitions. Though Evans (Liu and Evans, 1999) proves that the geometrical constraints can couple spatial fluctuations, the problem about coupling the local fluctuations at the infinite systems remains open. Indeed, the lack of boundary constraints and long-range interactions renders the local transitions non-correlated both in space and in time. This however, immediately introduces local strain, local overheating, sintering etc. At the next instant the spatial configuration changes and the local defects moves. Due time course their interaction can produce local reconstruction of the surface, can create mechanical defects etc. Eventually the process can yield either the system breakdown or the reaction termination.

There is a wide spectrum of works aimed to explore an effect of correlations of fluctuations in extended systems as an interplay among noise correlations, non-linearitity and spatial coupling. However, all the developed so far approaches to the task involve in the stochastic variables and noise sources a Wiener process whose increments are independent and unbounded. Thus, eventhough a cooperation of the fluctuations is available, the sequence of spatio-temporal configurations through which the it arrives to the global coupling varies in uncontrolded way that it is incompatible with the idea of the boundedness.

Hence, a transition between two states does not violate the stability of the system if and only if the rates of energy and/or matter exchange permanently do not exceed their local thresholds of stability. This constraint, called hereafter local boundedness, selects a set of accessible states to any given one. Obviously, the set of accessible states and the rates of energy and/or matter exchange is specific to the system and even to the state. Yet, the local boundedness is matched in the following common constraints of the state space: both the "distance" between any accessible states and the "velocity" of the motion anywhere in the attractor and at any instant should be finite. The latter requirement is in a sharp contrast with the infinite velocity obtained in the case when the transitions between any two states are allowed (Zaslavsky, 2002) and (Horsthemke, 1999).

Another important property of the state space is its global boundedness introduced by another necessary condition for keeping the evolution of any natural and artificial system stable, namely: the amount of energy and/or matter involved in the system is limited. So, the size of fluctuations should never exceed the macroscopic thresholds of stability of the system. The global boundedness renders the state space always confined in a finite volume. This allows to reenumerate the state space so that all accessible states to a given one are the nearest neighbors.

Thus, the relationship between the time development of fluctuations and the stability of the system introduces two new general properties of the state space and of its motion, namely: the state space is always bounded, the successive steps of motion are always finite and involve only nearest neighbours. It`s most important feature, proven in sec.1.2, is that it ensures a finite velocity of the motion in the state space.

Our task now is to account for the structure of a bounded state space subject to local boundedness and to answer the general question whether it has a stationary state and invariant measure.

We start with the question whether the state space is connected by a Markovian chain under the imposed constraints of the local and global boundedness. If so, we can apply the developed so far approaches such as Focker-Planck equation, master equation etc. to our problem. Otherwise, we should look for new approach(es) and to anticipate new properties.

Let us suppose that the transition from state $A$ to neighbor state $\beta_\theta$ depends only on whether the transition to $(0]$ has happened. However, the transition to $\Lambda$ depends on whether it comes from its nearest neighbourhood . Hence, the transition to $\beta^*$ is set on the chain of the previous transitions $N \gg m$. So, is our process non-Markovian though the Chapman-Kolmogorov relation holds?! On the one hand, it seems Markovian because the transition from $j$ to $i$ depends only on $j$. On the other hand, it is non-Markovian, because any admissible transition depends on the succession of the previous ones. (Examples of non-Markovian chains that fulfill Chapman-Kolmogorov relation are presented in (Feller, 1970)). The question now becomes whether the "non-Markovian" succession covers the whole history of the process. The clue is in the Lindedeberg theorem (Feller,1970) that states: any bounded irregular sequence (BIS) has finite mean and finite variance. Then, our trajectory "loses" its memory any time it crosses the mean. Though any "walk" that starts at the mean comes back to it in a finite time, the size of the "walks" varies up to the size of the state space. It means that a non-Markovian "walk" can comprise most of the states.

As a result, we should look for a new approach to establish whether our state space has an invariant measure.

At first, let us specify some features of the state space trajectories. The existence of more than one accessible state to any given state makes each transition rate a multi-valued function. Indeed, at any instant only one transition takes place. However, it is arbitrarily chosen among the transitions to all accessible states. This introduces certain stochasticity of the transition rate even when all its selections can be carried out explicitly. The outcome is that the trajectory appears as a kind of a fractal Brownian "walk". The stochastisity induced by the choice of one selection among all available breaks any possible long-range periodicity (i.e. the incremental memory has finite size). So, the trajectories in our state space are fractal Brownian walks subjects to an incremental boundedness, namely: finite size of the increments (i.e. transition rates), finite size of the incremental memory and arbitrary selection rules.

In turn, the boundedness of the incremental memory size distinguishes a scale above which all the scales up to the thresholds of stability contribute uniformly to the creation of the fluctuations. Then, it is to be expected that the coarse-grained structure of the attractor exhibits universal properties insensitive to the details of the incremental statistics. In particular, it is to be expected that these properties are the same in any direction of the attractor. So, we can study the one-dimensional sequence resoluted by the projection of a trajectory onto any direction. We should recall that every such sequence belongs to the class of bounded irregular sequences (BIS) subject to incremental boundedness defined above.

Indeed, in sec.1 we prove that any one-dimensional BIS subject to incremental boundedness has certain universal properties. It is found out that coarse-grained structure of each BIS is a succession of well separated one from another successive excursions. An excursion is a trajectory of a walk originating at the mean value of a given sequence at the moment $t$ and returning to it for the first time at the moment $t + \Delta$. The characteristics of each excursion are amplitude, duration and embedding interval. The latter implies that each excursion is loaded in a larger interval whose duration is interrelated with the duration of the excursion itself. The embedding does not allow any overlapping of the successive excursions and thus prevents the grow of the excursion amplitude to an arbitrary size. Thus, it preserves both the global and local boundedness arbitrarily long-term.

The incremental boundedness renders that each fluctuation has certain duration interrelated with its amplitude. By the use of this relation we prove in sec. 1.2 that the velocity of the motion on the state space trajectories is always finite. It is in accordance with our general requirement about the boundedness of the amount and rates of the energy and/or matter exchange involved in each transition.

An immediate consequence of the Lindeberg theorem is that every BIS is a subject to the Central Limit Theorem. So, the deviations from the mean are normally distributed. It should be stressed that this distribution is robust to the details of the "incremental" statistics, i.e. to the particularities of the transition rates and to the specific structure and properties of the sets of accessible states. Hence, the normal distribution appears as the target invariant measure of a bounded state space with bounded increments.

However, the Lindeberg theorem does not guarantee the asymptotic stability of the normal distribution as an invariant measure. The problem arises from the way in which the trajectories approach the boundaries of the state space, i.e. whether they "bump" into or "slide" past them. In the former case a motion termination is possible while in the latter case it proceeds. In other words, the question is whether there is mechanism that sustains the state space motion arbitrary long term confined to a finite state space.

The presence of the boundedness as the basic characteristic of our type state space suggests a parallel with the low-dimensional deterministic chaos - a phenomenon that occurs at the dynamics of simple deterministic systems. It is associated with unpredictability and great sensitivity to the initial conditions introduced by stretching and folding (Lichtenberg and Lieberman,1983) and (Zaslavsky,1983). However, the deterministic chaos also exhibits boundedness: the folding is provided by the fact that the dynamics of the discussed systems is confined to a finite volume of the state space. The stretching happens along the unstable directories and gives rise to the unpredictability.

In particular, our attention is focused on the folding because its presence sustains the evolution of a chaotic system permanently bounded to a finite phase volume. Intuitively it seems that the folding in our state space is ensured by the presence of the thresholds of the stability. However, it is to be expected that the particularities of the boundary conditions imposed by the thresholds of stability make the folding sensitive to them and thus not universal. The question now becomes whether there is a folding insensitive to the details of the boundary conditions. We consider this problem together with the issue about the folding viewed as a necessary condition for keeping the evolution confined in a bounded state space arbitrary long time. From this point of view, the folding is to be associated with the largest fluctuations (i.e. deviations from the mean), namely those whose amplitude is of the order of the thresholds of stability.

In the sec. 2 we find out that an universal folding do exists whenever certain relation among the thresholds of stability, a parameter set on the incremental statistics and the variance of the sequence holds. The derivation of the relation involves characteristics of the coarse-grained large-scaled fluctuations established in sec.1.

On the way of deriving the necessary condition for existing of a universal folding we prove that our attractor is a transitive dense set of periodic orbits and the value of the Lyapunov exponent is insensitive to the initial point of a trajectory. According to (Oseledec,1968) and (Ruelle,1989) these properties are representative for the deterministic chaos. However, along with these our attractor exhibits other chaotic properties as well. It has been proven (Koleva and Covachev,2001) that each BIS subject to an incremental boundedness has the following chaotic properties insensitive to the particularities of the incremental statistics:

(I) the power spectrum uniformly fits the shape $1/f^{\alpha(f)}$ where $\alpha(f) \to 1$ at $f \to 1/T$ ($T$ is the length of the sequence) and $\alpha(f)$ monotonically increases to $p > 2$ at $f \to \infty$.

(ii) the phase space attractor has non-integer correlation dimension $\nu(X)$ that monotonically decreases from $\nu(X) = d$ at the mean value to $\nu(X) = 0$ at the boundaries of the attractor; $d$ is the embedding (topological) dimension.

(iii) the Kolmogorov entropy is finite.

Thus, when the relation established in sec.2 holds, the coarse-grained structure of a bounded state space subject to incremental boundedness has an asymptotically stable invariant measure ( the normal distribution) and the motion in the state space exhibits chaotic properties. That is why we call this behavior "chaotic" kinetics.

It is worth noting that the boundedness appears naturally in the above considerations through the obvious physical constraint that each system has its finite local and global thresholds of stability associated with the amount and the rate of the energy and/or matter exchange.

However, when the evolution of an extended many-body system is developed under local rules, the lack of correlation among the local spatio-temporal fluctuations renders a self-sustained boundedness rather occasional and specific to a system or to certain states. This gives rise to the question about a mechanism that give rise to both local and global boundedness at each point of the state space.

Recently one of us has proposed a general mechanism for coupling spatio-temporal fluctuations at surface reactions that proceed at infinite interfaces (Koleva, 2002). The mechanism is insensitive to the particularities of the reaction mechanism, surface properties etc. As a result of the coupling, at any value of the control parameters the system exerts macroscopic fluctuations that are compatible with the local and global boundedness. The important property of this coupling mechanism is its persistence, i.e. for each state of the system it determines a non-zero set of states accessible from that one. In turn, this provides the "continuity" of the state space trajectories. Moreover, the distance between any given and each of the accessible states is always finite.

We give a brief presentation of that mechanism in sec.3. The aim is to elucidate those its general properties that make the macroscopic evolution of the surface reactions to be the subject of the "chaotic" kinetics.

In sec. 4 we present several experimental evidences that for the chaotic kinetics.

## 1.Properties of the Large-Scaled Fluctuations

The task of the present section is to work out those properties of the BIS`es by the use of which we can prove the existence of a universal folding. In turn, it ensures the asymptotic stability of the invariant measure (normal distribution) of the attractor.

Let us recall that the boundedness of the incremental memory size distinguishes a scale above which all the scales up to the thresholds of stability contribute uniformly to the creation of the fluctuations. Thus, it is to be expected that the coarse-grained properties of the large-scaled fluctuations are insensitive to the particularities of the incremental statistics. It gives rise to the universality not only of the folding but to the universality of the coarse-grained structure of the attractor.

In the next subsection we prove that the coarse-grained large scale fluctuations of each BIS appear as a sequence of separated by non-zero intervals successive excursions. It should be stressed that this structure is a result of merely the global and the incremental boundedness.

Each excursion is characterised by its amplitude, duration and embedding interval. The separation of the successive excursions means embedding of each of them in a larger interval so that no other excursions can be found in that interval. The duration of the "embedding" interval is a multi-valued function whose properties are strongly related to the duration of the embedded excursion $\Delta$ itself: the range and the values of of the selection are set on $\Delta$; the realisation of any excursion is always associated with the realisation of its embedding interval. Since the duration of the each embedding interval is a multi-valued function, its successive performances permanently introduce stochasticity through the random choice of one selection among all available. Thus, this induced stochastisity breaks any possible long-range periodicity ( i.e. large-size memory) and helps the excursion sequence to preserve the chaotic properties listed in the Introduction.

The relation $\Delta \leftrightarrow A$ is set on a "blob" structure of the incremental walk that produces a BIS, namely: because of the finite size of the incremental memory, on coarse-grained scale any fractal Brownian walk can be considered as a symmetric random walk of "blobs" created by subwalks whose size counterparts the size of the incremental memory.

Our first task is to work out explicitly the relations $T \leftrightarrow \Delta \leftrightarrow A$, i.e. the relations between the duration of the embedding time interval $T$, respectively the duration $\Delta$ and the amplitude of the corresponding excursion $A$.

## 1.1 Embedding time intervals. Relation $T \leftrightarrow \Delta$

The major role of the "embedding" is that it does not allow the overlapping of the successive excursions and thus prevents the grow of the excursion amplitude to an arbitrary size. So, the "embedding" preserves the global boundedness arbitrarily long-time.

The present task is to work out explicitly the relation between the duration of the embedding intervals and the duration of the corresponding excursions. That relation is based on the notion of an excursion: a trajectory of a walk originating at the mean value of a given sequence at time $t$ and returning to it for the first time at time $t + \Delta$. Therefore, the probability for an excursion of duration $\Delta$ is determined by the degree of correlation between any two points of the sequence. On the other hand, the probability that any two points of a sequence, separated by distance $\eta$, have the same value is given by the autocorrelation function $G(\eta)$. A generic property of the BIS`es (Koleva and Covachev, 2001) is that the coarse-grained autocorrelation function of any of them can be defined for sequences of arbitrary but finite length $T$. Yet, its shape is universal, namely:

$$G(\eta, T) \propto 1 - \left(\frac{\eta}{T}\right)^{\nu(\eta/T)}. \tag{1}$$

where $\nu\left(\frac{\eta}{T}\right)$ is an continuous everywhere monotonically decreasing between the following limits function:

$$\nu\left(\frac{\eta}{T}\right) \to p - 1 \quad \text{as} \quad \frac{\eta}{T} \to 0 \tag{2a}$$

$$\nu\left(\frac{\eta}{T}\right) \to 0 \quad \text{as} \quad \frac{\eta}{T} \to 1. \tag{2}$$

Then, the probability that an excursion of duration $\Delta$ happens in an interval $T$ reads:

$$J = \int_a^b x^{\pm\alpha(x)} dx = \frac{b^{\alpha(b)+1}}{1\pm\alpha(b)} - \frac{a^{\alpha(a)+1}}{1\pm\alpha(a)} \ . \tag{3}$$

Thus,

$$P(\Delta,T) = \frac{1}{T}\int_0^\Delta G(\eta,T)d\eta = P(\Delta,T) = \int_0^{\Delta/T}\left(1-\varepsilon^{\nu(\varepsilon)}\right)d\varepsilon \tag{4}$$

By the use of the standard calculus (Koleva, 2002), the integration of any power function of non-constant exponent $\alpha(x)$ ($\alpha(x)$ is an everywhere continuous function) reads

$$P(\Delta,T) = \frac{\Delta}{T}\left(1-\left(\frac{\Delta}{T}\right)^{\nu\left(\frac{\Delta}{T}\right)}\right). \tag{5}$$

The $P(\Delta,T)$ dependence only on the ratio $\Delta/T$ verifies the assumption that every excursion of duration $\Delta$ is "embedded" in an interval of duration $T$ so that no other excursion happens in that interval.

    The next step is to work out the shape of $P(\Delta,T)$. Its role is crucial for the behavior of the excursion sequences. To elucidate this point let us consider the two extreme cases:

    (I) $P(\Delta,T)$ is a sharp single-peaked function. It ensures a single value of the most probable ratio $\frac{\Delta}{T}$. So, when the sequence involves identical excursions, their sequence matches a periodic behavior. The latter, however, introduces long-range correlations in the incremental statistics.

    (ii) $P(\Delta,T)$ has a gently sloping maximum. Then, the relation between $\Delta$ and $T$ behaves as a multi-value function: a range of nearly equiprobable values of $T$ corresponds to each most probable $\Delta$. Thus, this induced stochastisity breaks any long-range periodicity (i.e. large-size memory among increments).

    The establishing of the particular shape of $P(\Delta,T)$ requires the knowledge about the explicit shape of $\nu\left(\frac{\Delta}{T}\right)$. Next it is figured out on the grounds of the proof that neither BIS sustained to an arbitrary length comprises any long-ranged increment correlations. The general restriction on the increment correlation size requires an uniform contribution to the power spectrum of all scales, i.e. there are no "special" frequencies at the power spectrum.

    The only factor that can modify the power spectrum is the non-constant exponent $\alpha(f)$ of its shape $1/f^{\alpha(f)}$. The boundedness requires the monotonic decay of $\alpha(f)$ in the limits $[1,p]$ not specifying its shape (Koleva and Covachev, 2001). Our task now is to establish the shape(s) of $\alpha(f)$ that fits the lack of long-range increment correlations. In virtue of the strict monotony of the power spectrum the required criterion is that neither any its component nor any its derivative of arbitrary order has a specific contribution. Simple calculations yield that it is achieved if and only if the shape of $\alpha(f)$ is the linear decay, namely:

$$\alpha(f) = (1+\gamma f) \tag{6}$$

Eq. (6) provides that the $1/f^{\alpha(f)}$ derivative of an arbitrary order has the same sign throughout the entire frequency interval of the power spectrum. On the contrary, for any non-linear decay of $\alpha(f)$ each order derivative of $1/f^{\alpha(f)}$ changes its sign at certain frequencies.

Thus, only the linear decay does not introduce any additional scale to those inherent for the increment statistics.

Because of the diffeomorfism between $\alpha(f)$ and $\nu\left(\frac{\Delta}{T}\right)$ it is obvious that the shape of $\nu\left(\frac{\Delta}{T}\right)$ reads:

$$\nu\left(\frac{\Delta}{T}\right) = (p-1)\left(1 - \frac{\Delta}{T}\right). \tag{7}$$

The plot of $P(\Delta, T)$ with the above shape of $\nu\left(\frac{\Delta}{T}\right)$ shows that it has a gently sloping maximum: indeed, the values of $P(\Delta, T)$ in the range $\frac{\Delta}{T} \in [0.25, 0.4]$ vary by less than 7%. Outside this range $P(\Delta, T)$ decays sharply. Thus, though $P(\Delta, T)$ is a single-valued function, it provides a multi-valued relation between the most probable values of $\Delta$ and $T$, namely: a certain range of nearly equiprobable values of $T$ is associated with each $\Delta$. In the course of the time the multi-valued relation is exerted as a random choice of the duration of the "embedding" intervals even when the sequence comprises identical excursions.

### 2.2. Relation $A \leftrightarrow \Delta$. Symmetric Random Walk as the Global Attractor for the Fractal Brownian Motion. Finite Velocity

The general constraint of the incremental boundedness gives rise to the expectation that there is certain uniform relation between the amplitude of each excursion and its duration. The relation amplitude $\leftrightarrow$ duration determines not only a property of an excursion but the "velocity" of the "motion" on that excursion. It has already been established in the Introduction that each state space trajectory can be considered as a fractal Brownian walk. So, being a one-dimensional projection of a state space trajectory, any presently considered BIS is a fractal Brownian walk as well. The latter provides certain relation between the amplitude $A$ and the duration $\Delta$ of an excursion, namely: $\sqrt{\langle A^2 \rangle} \propto \Delta^{\beta(\Delta)}$, $\beta$ is set by the particularity of the increment statistics; the averaging is over the sample realisations. The dependence of $\beta$ on $\Delta$ comes from the interplay between the finite radius of the increment correlations $a$ and the amplitude of the excursion itself that is limited only by the thresholds of stability.

Because of the finite size of the incremental memory, on coarse-grained scale any fractal Brownian walk can be considered as a symmetric random walk of "blobs" created by subwalks whose size is a counterpart of the incremental memory size. The implication that the incremental memory has a finite size renders that the incremental walks that create blobs has a finite length $m$; the particularities of the incremental statistics determines an exponent $\beta$, such that the $\sqrt{m.s.d.}$ of the blob creating walks equals $m^\beta$. Then, the large excursions can be approximated by a symmetric random walk with a constant step equal to the size of the blobs. Thus, the dependence of any large scale excursions on its duration reads:

$$\sqrt{\langle A^2 \rangle} \propto N^{0.5} m^\beta \tag{8}$$

where $N$ is the number of the blobs.
It is obvious that when $N \gg m$ the dependence tends to:

$$\sqrt{\langle A^2 \rangle} \propto N^{0.5} a \qquad (9)$$

where is considered constant independent of $N$. So, the symmetric random walk with constant step appears as the global attractor for any fractal Brownian motion.

An important outcome of eq.(8) is that it sets the velocity of the motion on the excursion. It is obvious that this relation provides a finite velocity whenever there is diffeomorfism between $A$ and $\Delta$. Evidently, it is provided by any non-zero but finite value of $\beta$. The finite incremental boundedness is manifested in a non-zero value of $\beta$. In addition, the fractal Brownian walk derivation of $\beta$ implies that the latter is bounded in the range $(0,1]$. Any non-zero but finite value of $\beta$ makes the size, duration and the structure of the blobs to depend strongly on the incremental statistics and thus to be specific to the BIS. To compare, suppose $\beta = 0$. In this case, the size and the duration of the blobs are independent from one another and appear as parameters.

Another outcome of our considerations is that the finite size of the "blobs" ensures the uniform convergence of the average to the mean of the original BIS. Indeed, the distinctive property of any fractal Brownian motion is that any exponent $\beta \neq 0.5$ arises from an arbitrary correlation between the current increment $\mu_i$ and the corresponding step $\tau_i$. Then the average $\overline{A}$ reads:

$$\overline{A} = \sum_{i=1}^{N} \mu_i(\tau_i)\tau_i = \sum_{i=1}^{N} (-1)^{\gamma_i} \tau_i^{\beta_i} \qquad (10)$$

and correspondingly the m.s.d. :

$$\langle A^2 \rangle \propto \left\langle \sum_{i=1}^{N} (\mu_i(\tau_i)\tau_i)^2 \right\rangle = \left\langle \sum_{i=1}^{N} \tau_i^{2\beta_i} \right\rangle \qquad (11)$$

where the averaging is over the different samples of the trajectory; The property of the above relations is that whenever the probabilities for $\gamma_i$ odd and even are not permanently equal there is a correlation between the increment and the corresponding step. So $\overline{A}$ is certainly non-zero that immediately makes that the deviation from the mean non-zero. Moreover, eq.(10) yields that $\overline{A}$ can become arbitrarily large on increasing $N$. On the contrary, a permanent equal probability for $\gamma_i$ odd and even means independence from one another of the increments and the steps. It yields $\overline{A} = 0$ which guarantees the uniform convergence of the average to the mean.

A major property of any BIS is the existence of mean and variance that is guaranteed by the Lindeberg theorem (Feller,1970). This brings about two very important consequences. The first one is that the amplitudes of the excursions are normally distributed. The second one is that the excursion sequence is a stationary process. Indeed, the boundedness and the finite-size memory render a uniform convergence of the average to the mean of every BIS. In turn, it provides the stationarity of the excursion. Then, the frequency of occurrence of an excursion of a size $A$ is time-independent and reads:

$$P(A) = cA^{1/\beta(A)} \frac{\exp(-A^2/\sigma^2)}{\sigma} \qquad (12)$$

The required probability $P(A)$ is given the duration $\Delta = A^{1/\beta(A)}$ of an excursion of amplitude $A$ weighted by the probability for appearance of excursion of that size (normal distribution). $\sigma$ is the variance of the BIS and in the present consideration is a parameter; $c = \dfrac{1}{\sigma^{1/\beta(\sigma)}}$ is the normalizing term. The stationarity of the excursion appearance ensures that $P(A)$ has the same value at every point of the sequence.

The behavior of any BIS is inherently related to the incremental statistics trough the explicit dependence of $P(A)$ on $\beta(A)$ in eq.(12). However, when the size of the required excursion is much larger than the size of the symmetric random walk, $\beta(A)$ turns constant equal to $0.5$. Then, $P(A)$ gradually gets insensitive to the details of the incremental statistics. Thus, whenever $A >> a$ the behavior of the excursions becomes totally insensitive to the incremental statistics.

## 2. Universal Folding Mechanism of the BIS`es

The goal of this section is to establish the necessary conditions for the asymptotic stability of the invariant measure (the normal distribution) of an attractor with global and incremental boundedness. The global boundedness suggests that the motion on such attractor creates a kind of a stretching and folding. Then, the asymptotic stability of the invariant measure is to be associated with folding that is insensitive to the way how a trajectory approaches the thresholds of stability. Thus, our aim is to find out the conditions under which such insensitive to the "boundaries" folding exists.

From the viewpoint of the deterministic chaos the folding is associated with a negative value of the Lyapunov exponent. As a measure of unpredictability the latter is the average measure how fast a trajectory deviates under arbitrary small perturbation of the initial conditions. On the other hand, from the point of view of a BIS`es, it is to be associated with the largest fluctuations, namely those whose amplitude is of the order of the thresholds of stability. Thus, our first task is to define the Lyapunov exponent in terms of the large scaled fluctuations and to show explicitly the dependence of its value and sign on their characteristics.

A necessary step for the universality of the folding is the insensitivity of the large-scaled fluctuations to the details of the incremental statistics. It has been already established that every coarse-grained BIS has 3 specific parameters: values of the threshold of stability $A_{tr}$, the variance $\sigma$ and the power $\beta(A)$ in the relation amplitude$\leftrightarrow$duration of the excursions. Next we prove that certain relation among these parameters provides the existence of the target folding.

From the viewpoint of the deterministic chaos the folding is associated with a negative value of the Lyapunov exponent whose rigorous definition reads:

$$\xi = \lim_{t \to \infty} \frac{1}{t} \ln |U(t)| \qquad (13a)$$

where

$$U(t) = X(t) - X^*(t) \qquad (13b)$$

$X^*(t)$ is unperturbed trajectory and $U(t)$ is the average deviation from it. So $|U(t)|$ is the measure of all the available deviations from a given point $X^*$.

On the other hand, from the point of view of BIS`es, the Lyapunov exponent is to be associated with the large-scaled fluctuations since they give rise to the essential deviations from the mean.

Now we are ready to write down the asymptotic explicit expression for the average deviation from a trajectory that starts at $X^*$. The corresponding $|U(t)|$ set on the terms of the excursions reads:

$$|U(t)| = \int_{A^*}^{A_{tr}} AP(A)dA + \int_{A_{cgr}}^{A^*} AP(A)dA \qquad (14)$$

$A_{cgr}$ is the level of coarse-graining, i.e. averaging over all scales smaller than $A_{cgr}$. This "smoothes out" all the excursions whose sizes are smaller than $A_{cgr}$ and renders their contribution to the Layapunov exponent zero.

The separation into two terms each of which represents the deviations from $A^*$ to larger and smaller amplitudes is formal. It is made only to elucidate the idea that starting at any point of the attractor one can reach any other through a sequence of excursions. Thus, the Lyapunov exponent $\xi$ reads:

$$\xi = \ln \int_{A_{cgr}}^{A_{tr}} AP(A)dA \tag{15}$$

Here we come to the same result as the Oseledec theorem states (Oseledec, 1968), namely: the Lyapunov exponent for the chaotic systems does not depend on the initial point of the trajectory.

On the other hand, the stationarity of the excursion process makes the set of excursion sequence a dense set of periodic orbits. Moreover, it renders its transitivity as well: starting anywhere in the attractor a sequence of excursions can reach any other point in it. Some authors (Ruelle, 1989) list these properties as a definition of the chaos. Here they appear as a result of the global and the incremental boundedness of the BIS`es. It is worth noting that the chaotic properties listed in the Introduction are also a result of the global and the incremental boundedness. It supports the suggestion about the paramount role of the boundedness at defining the chaoticity.

It should be stresses that the chaotic properties listed in the Introduction are derived under the condition that all scales larger than the blob size $a$ contribute uniformly to the stochastic properties of any BIS. The above chaotic properties also do not involve any specific scale larger than the blob size. However, it seems that it brings a contradiction: how the scale-free process of excursion sequence interferes with the boundary conditions imposed by the presence of the thresholds of stability. The contradiction is solved by the presence of folding since the latter makes the approach to a boundary a tangent turn back. Thus, the folding being a necessary condition for keeping the evolution of a BIS permanently confined in a finite attractor ensures that the chaoticity produced by the stretching and folding is a scale-free process.

It is to be expected that the size of an excursion determines its contribution to the stretching or folding of a trajectory. Indeed, the small size excursions are random walks whose frequency is essentially high (eq.(12)). Figuratively speaking, they "hold" any trajectory permanently deviated from the mean. So, the small size excursions most probably contribute to the stretching of the trajectories. On the contrary, the largest excursions are rather occasional and any trajectory subjected to them spends most of its time closest to the mean. So, they would provide the folding. The explicit revealing of the role of small and large excursions is made by the use of the coarse-graining: the role of the excursion size is carried out by scanning the ratio $A_{cgr}/\sigma$.

The ratio $A_{cgr}/\sigma$ has two extreme cases:

(I) $\dfrac{A_{cgr}}{\sigma} \ll 1$, i.e. the contribution of the small excursions prevails. By the use of the steepest descent method, eq.(15) yields:

$$\xi \approx \ln \sigma \tag{16}$$

Eq.(16) tells that asymptotically any trajectory visits any point of the attractor so that the mean deviation from any initial point is the same for every trajectory and is bounded by the thresholds of the attractor itself. This makes the value of $\xi$ positive. The latter justifies our speculation that the small size excursions contribute to the stretching. Further, visiting of any point of the attractor starting anywhere in it makes the motion on our attractor ergodic.

(ii) $A_{cgr} \gg \sigma$, i.e. large scale excursions contribution prevails. Eq.(15) yields:

$$\xi \approx \left(\frac{1}{\beta(A_{cgr})}+1\right)\ln\frac{A_{cgr}}{\sigma}+\ln\sigma-\frac{A_{cgr}^2}{\sigma^2} \qquad (17)$$

While $\xi$ from eq.(16) is always positive which provides stretching, eq.(17) opens the alternative for $\xi$ being both positive or negative setting on the relation among $\beta(A_{cgr}), \sigma$ and $A_{cgr}$. The permanent presence of folding is necessary for keeping the evolution bounded in a finite size attractor arbitrary long time. The natural measure of the folding is the negative value of the Lyapunov exponent. Thus we come to the condition: the largest size fluctuations provide folding whenever $A_{tr}$, $\beta(A_{tr})$ and $\sigma$ are such that $\xi < 0$. So our target relation reads:

$$\xi \approx \left(\frac{1}{\beta(A_{tr})}+1\right)\ln\frac{A_{tr}}{\sigma}+\ln\sigma-\frac{A_{tr}^2}{\sigma^2} < 0 \qquad (18)$$

It is worth noting that the eq.(18) provides that the excursions "turn back" without "feeling" the boundaries of the attractor. This makes the folding insensitive to the details of the thresholds of stability and the way they are approached.

It is worth noting that the folding is a broader notion than a tangent approach to a boundary. Both the folding and the tangent approach produce the same effect: they contribute to the convergence of a trajectory making it to depart from the threshold. Yet, the tangent approach itself is a property of the random walk that creates the excursions along with appropriate boundary conditions imposed, while the folding is provided by eq.(18) not involving any boundary conditions.

## 3. Diffusion-Induced Noise, Synchronization and Evolutionary Equations

The goal of the present section is to present an example that meets the criteria for both local and global boundedness, namely: the accessible transitions are at the neighborhood of any given state, the incremental memory has finite size, the state space is bounded.

Yet, the focus is put on the physical mechanism that endure the boundedness arbitrarily long-time and throughout the entire state space. It is presented in the first subsection.

The second subsection elucidates certain properties that are explicitly introduced by the boundedness and on the other hand are generic for all the surface reactions that proceed at the extended interfaces gas/solid.

### *3.1. Diffusion-Induced Noise and Synchronization*

Next we consider the surface reactions that proceed at infinite interfaces gas/solid exposed to steady external constraints. Recently one of us (Koleva 1998), (Koleva and Covachev, 2001) put forth a driving mechanism of a local quantum phenomenon, called diffusion-induced noise, that gives rise to rather unusual macroscopic effects. Its extraordinarity is founded on an apparent association with the stability of the system. Thus, the modelling of this phenomenon requires a new approach that explicitly involves the issue of stability. The importance of the diffusion-induced noise is that it is inevitable and generic for all the surface reactions and its presence and certain properties are insensitive to the particularities of the gas-solid system.

To elucidate the ubiquity of the considered phenomenon we present the driving mechanism of the diffusion-induced noise for the adsorption, since it is a step prerequisite of any surface reaction at any control parameter choice. The mechanism is based on the interplay between: (I) the lack of correlation between moments and points of the gas phase species trapping at the surface; (ii) the generic property of the most adlayers that no more than one species can be adsorbed at a single active site. That interplay causes fundamental changes of the properties of the overall probability for adsorption (correspondingly the adsorption rate). Given is a species trapped in a vacant site. Its further relaxation to the ground state can be interrupted by an adspecies that arrives at the same site and most probably occupies it. Thus the adspecies violates the further trapped species relaxation at that site, since no more than one species can be adsorbed at a single site. The trapped species can complete the adsorption if and only if after migration it finds another vacant site. The impact of the adspecies intervention to the trapped species probability for adsorption is twofold: first, it cannot be considered as a perturbation, since it changes the adsorption potential qualitatively, namely from attractive it becomes repulsive. That is why, that type of interaction has been called diffusion-induced non-perturbative interaction. Second, the lack of coherence between the trapping moment and the moment of adspecies arrival makes the probability for adsorption multi-valued function: each selection corresponds to a certain level of relaxation at which a diffusion-induced non-perturbative interaction happens. Therefore, the adspecies mobility brings about a fundamental duality of the probability for adsorption (and of the adsorption rate correspondingly): though each selection can be computed at microlevel, the establishing of a given selection is a stochastic process since it is a random choice of a single selection among all available.

It is obvious that the presence of the diffusion-induced noise is insensitive to the details of the adsorption Hamiltonian and thus to the particularities of the adspecies and the interface. Further, it is presented at each value of the control parameters. So, the diffusion-induced noise persists at each point of the state space and is generic for any surface reaction.

Further, since the diffusion-induced non-perturbative interactions are local events, the non-correlated mobility of the adspecies produces a lack of correlation between the established selections at any distance and at any instant. As a result, the produced adlayer would be always spatially non-homogeneous even in the academic case of identical adsorption and mobility properties of all types of adspecies. Outlining, the non-correlated diffusion-induced non-perturbative interactions always make the adsorption rates that come from different adsorption sites non-identical that immediately yields spatial non-homogeneity. Furthermore, the latter would be permanently sustained by the lack of coherence between the trapping moments and the adspecies mobility. In turn, the adlayer configuration would vary in an uncontrolled way which in a short time would cause either the reaction termination or the system breakdown. Thus, a stable long-term evolution is available if and only if there is a mechanism that suppresses the induced non-homogeneity. The importance of the problem is rendered by the fact that the latter should be ubiquitous for all the surface reactions at each parameter choice.

The only way out is to look for a mechanism that removes the induced non-homogeneity through a feedback that couples the local fluctuations. The supposition about a mechanism of fluctuations "dying down" leads to nowhere because there is permanent producing of local fluctuations - the gas phase bombardment of the surface is enduring and insensitive to what happens there.

Next it is supposed that the physical interactions among adspecies are only short range ones. Thus, the only way the excited species can interact is the collisions brought about by their mobility. It is worth noting that these collisions act towards evening of the current states of the colliding species. However, evening of the current states does not even the initially non-identical adsorption rates.

A successful coupling mechanism needs a feedback that acts toward evening of the initially non-identical adsorption rates through making the coupled species "response" to further perturbation(s) coherent. It is founded on a strong coupling adlayer-lattice, namely: the energy of colliding species dissipates to local cooperative excitations of the lattice. In turn, the impact of these local excitations on the colliding species is supposed large enough to induce a new transition that dissipates through the excitation of another local cooperative excitation etc. The feedback ceases its action whenever the colliding species response becomes coherent.

To make the above feedback available we have introduced two major assumptions (Koleva, 2002). The first one is that any adsorption Hamiltonian separates into a "rigid" and a "flexible" part under the disturbances induced by the non-correlated mobility of the adspecies. The "rigid" part is specific to the system and comprises the low-lying levels where the adspecies impact can be treated as perturbation. The "flexible" part corresponds to that highly excited states where the impact of the mobile adspecies is as strong as potentials that create the Hamiltonian. So, the "flexibility" of the Hamiltonian means that it always changes under the non-correlated movements. The sensitivity to the corresponding environments renders the "flexible" parts of the Hamiltonians that comes from different adsorption sites non-identical. Next, any change of a given Hamiltonian changes the state of the corresponding trapped (highly excited) species. Thus, the "flexibility" induces permanent transitions of any trapped species.

The next major assumption introduced in (Koleva, 2002) is about the general properties of the induced transitions. It is supposed that these transitions are non-radiative and dissipate always through excitation of appropriate local low-frequency co-operative excitations, e.g. acoustic phonons. The new point is that the local low-frequency excitations participate to the "flexible" Hamiltonian the same way as the mobile adspecies. This, in turn constitutes a non-perturbative feedback, namely: the mobile environment of any trapped species induces a non-radiative transition that dissipate through excitation of local acoustic phonon(s) (or other appropriate excitation(s)). The latter causes an immediate non-perturbative change of the "flexible" part of the Hamiltonian which in turn induces a new transition.. The feedback acts toward evening of the initially non-identical adsorption rates via "synchronization" of the initially non-identical "flexible" Hamiltonians so that they "respond" coherently to further perturbation(s). After the synchronisation is completed, the further relaxation proceeds through the "rigid" part of the Hamiltonian.

Further, it has been proven that though the "flexible" parts of the Hamiltonians are not identical, they have chaotic properties so that, following Berry, (Berry, 1985) their spectra share the same distribution of the nearest level spacing, namely the Wigner distribution. This helps the insensitivity of the feedback of the particularities of the adspecies since the Wigner distribution involves a single parameter to characterize the chaotic states and the transition between them.

Next, it is shown that the collisions between the species in chaotic states, i.e. the weakest perturbations that drive the feedback, dissipate through the excitation of acoustic phonons. The particular property of the feedback based on dissipation through acoustic phonons is that the coupling area (the wave-length) enlarges with a decrease of the energy of a transition that drives the feedback (frequency).

Further, it is proven that the global adsorption rate does not depend on the details of the spatio-temporal configuration of the "chaotic" species. To the most surprise, it is always identical to the individual adsorption rate that comes from certain local configuration. It is verified that the feedback always selects that individual adsoprtion rate which initially is in the most favorable local configuration: such that the difference in the state of that species and its immediate neighbors is the smallest. It is proven as well that the synchronisation is a scale-free process that does not blur the individual properties of the "most favourable" adsorption rate.

The most pronounced differences among the individual adsorption rates come from the undergoing of a diffusion-induced non-perturbative interaction. Thus the particularity of the "most favourable" adsorption rate is determined by the established selection. Furthermore, the lack of correlation between the trapping moments and the adspecies mobility makes all the selections equiprobable. Therefore, at next instant the "most favorable" adsorption rate involves another selection of a diffusion-induced non-perturbative interaction. Consequently, the random choice of a single selection among all available brings about irregular variations of the global adsorption rate in the course of time.

The hallmarks of the variations of the adsorption rate are their obvious boundedness, persistence at any point of the state space and the duality determinsm-stochastisity. The latter implies that though each selection can be computed by an appropriated quantum approach, the establishing of a single selection is a stochastic process since it is random choice of a single selection among all available. However, the most prominent property of these variations is that they come from a coupling process. Thus, the current variation is the same for each of the adspecies.

*3.2. Evolutionary equations*

The permanent bounded variations of the adsrption and reaction rates give rise to rather unusual properties of the temporal behavior of the surface reactions that proceed at infinite interfaces gas/solids. Yet, some of these properties are generic for the macroscopic evolution of the latters. At first, the type of equations of the macroscopic evolution changes their type, namely from systems of autonomous ordinary or partial differential equations they become non-autonomous because of the permanent variations of the adsorption and reaction rates. Next we consider certain general properties introduced by the change of the type of the equations for case of ordinary differential equations. At steady external constraints the equations of the macroscopic evolution read:

$$\frac{d\vec{X}}{dt} = \vec{\alpha}\hat{A}(\vec{X}) - \vec{\beta}\hat{R}(\vec{X}) \tag{19}$$

where $\vec{\alpha}$ and $\vec{\beta}$ are these parts of the rates that explicitly depend on the external constraints and thus, for steady conditions, appear as control parameters in eqs.(19); $\vec{X}$ is the vector of the reaction species concentrations; $\hat{A}(\vec{X})$ and $\hat{R}(\vec{X})$ are the matrices of the adsorption and reaction rates of the reactants and the intermediates. The r.h.s. of eq.(19) matches the reaction mechanism. It should be stressed that at any instant each component of $\hat{A}(\vec{X})$ and $\hat{R}(\vec{X})$ is a selection randomly chosen among all available.

To elucidate the role of the permanent variations of the adsorption and reaction rates let us rewrite eq.(19) in the form:

$$\frac{d\vec{X}}{dt} = \vec{\alpha}\hat{A}_{av}(\vec{X}) - \vec{\beta}\hat{R}_{av}(\vec{X}) + \vec{\alpha}\hat{\mu}_{ai}(\vec{X}) - \vec{\beta}\hat{\mu}_{ri}(\vec{X}), \tag{20}$$

where $\hat{A}_{av}(\vec{X})$ and $\hat{R}_{av}(\vec{X})$ are the mean values of the adsorption and reaction rates at given parameter choice $\vec{\alpha}$ and $\vec{\beta}$; $\hat{\mu}_{ai}(\vec{X}) = \hat{A}(\hat{X}) - \hat{A}_{av}(\vec{X})$ and $\hat{\mu}_{ri}(\vec{X}) = \hat{R}(\vec{X}) - \hat{R}_{av}(\vec{X})$. Then, in the course of the time $\hat{\mu}_{ai}(\vec{X})$ and $\hat{\mu}_{ri}(\vec{X})$ appear as successive terms of corresponding zero-mean bounded irregular sequences. The subscript $i$ serves to stress that only one selection, randomly chosen among all available, takes place at a given instant. $\hat{\mu}_{ai}(\vec{X})$ and $\hat{\mu}_{ri}(\vec{X})$ has Markovian property in the sense that the probability for occurrence of a given selection does not depend on the probability for the appearance of any selection at the previous instant.

A representative property of the eqs.(20) has been worked out in (Koleva and Covachev, 2001). It has been established that at any parameter choice the power spectrum of the solution of eq.(20) comprises additively two parts. The first one is the power spectrum of $\vec{X}_{av}$ settled by the following equations:

$$\frac{d\vec{X}_{av}}{dt} = \vec{\alpha}\hat{A}_{av}(\vec{X}_{av}) - \vec{\beta}\hat{R}_{av}(\vec{X}_{av}) \tag{21}$$

Eqs.(21) is a system of autonomous ordinary differential equations, non-linear in general, and thus its solution can vary with the parameter choice, i.e. it is not limited to a single steady state only but e.g. it can be a limit cycle.

The second part of the power spectrum of $\vec{X}(t)$ is the power spectrum of the zero-mean BIS resoluted by the variations of $(\vec{X}(t) - \vec{X}_{av})$. It has been proven that the power spectrum of any zero-mean BIS with bounded incremental statistics is a continuous band that fits the universal shape $1/f^{\alpha(f)}$, where $\alpha(f) \to 1$ at $f \to 1/T$ ($T$ is the length of the sequence) and $\alpha(f)$ monotonically increases to $p > 2$ at $f \to \infty$. This shape is insensitive to the details of the incremental statistics. In our case, this implies that the shape $1/f^{\alpha(f)}$ is robust to the particularities of the adsorption and reaction mechanism involved in $\hat{\mu}_{ai}(\vec{X})$ and $\hat{\mu}_{ri}(\vec{X})$.

To outline, the separation of the power spectrum into two parts is robust to the particularities of r.h.s. of eqs.(20). Thus, in particular, it is robust to the details of the reaction mechanism and is ubiquitous for the surface reactions.

It should be stressed that the power spectrum of the solution of neither system of ordinary differential equations (eqs.(21)) can comprise simultaneously a discrete and a continuous band in its power spectrum. So, the coexistence of a discrete band, coming e.g. from a limit cycle, and a continuous band of the above shape serves as an evidence that the macroscopic evolution of a system is a described by the equations of the type given by eqs.(20).

Another effect caused by the stochastic terms in eqs.(20) are the permanent deviations from the dynamical regime prescribed by eqs.(21) even when the control parameters are physically kept fixed. This is so because any difference $(\vec{X}(t) - \vec{X}_{av})$ can effectively be presented as a solution of eqs.(21) at "shifted" control parameters. The latter immediately causes a change either of the characteristics of the original dynamical regime or it even induces a bifurcation. It is obvious that at a given parameter choice an induced bifurcation needs a fluctuation of certain size. On the other hand, the sequence resoluted by $(\vec{X}(t) - \vec{X}_{av})$ is a zero-mean BIS with bounded incremental statistics. So, the properties of the large-scaled fluctuations match those already established in sec.1. Two of these properties, the limited duration of any large scaled fluctuation and the embedding make each induced bifurcation to have a temporal effect - it appears at some instant and disappears at another one.

Other general properties of the induced bifurcations are: at each value of the control parameters an induced bifurcation needs development of a fluctuation of certain size. The corresponding power spectrum always comprises a continuous band of the shape $1/f^{\alpha(f)}$. As established in sec.1.2, on increasing of their size, the development of the large fluctuations becomes less and less sensitive to the details of the reaction mechanism.

The above properties make the induced bifurcations drastically different from those considered so far. Indeed, so far it has been supposed that the noise is important only near bifurcation points (Nicolois and Prigogine, 1977). Here we come to the result that the diffusion-induced noise can generate temporal bifurcations even when the control parameters are far from a bifurcation point.

## 4. Comparison to the Experiment

The goal of the present section is to show an example that manifests the typical properties of a macroscopic evolution established in the previous subsection.

We have studied (Koleva et al, 2000) the reaction system $HCOOH + O_2 + N_2$ supported over $Pd$ catalyst (2 different charges of Heraeus gas purification catalyst containing 0.5 $wt\%$ $Pd$ on $Al_2O_3 - SiO_2$). Its temporal behavior has been studied in the temperature interval $110 - 130\ °C$ by means of Kral type continuous flow calorimeter. The $O_2$ feed concentration was varied from $0.5 - 12\%$ while that of $HCOOH$ - from $0.5\%$ to $8\%$. The contact time $\tau$ was varied from 0.3 to 13.6 $sec$. The reactor was charged with 10 to 20 $mg$ of ground catalyst ($63 - 100$ $\mu m$ sieve fraction).

The total $BET$ surface area of the first charge ($No.68323/1977$) amounts to 380 $m^2/g$, the $Pd$ metal surface area being 0.5 $m^2/g$. Approximating the $Pd$ particles on the catalyst surface as cubes, the edge length has been calculated to be $a = 42\ Å$ while the $X-ray$ measurements gave $a < 50\ Å$ for the particle size. The size has also been estimated by means of transmission electron microscope - it varies from 12 to 50 $Å$. The respective characteristics of the second charge ($No.83247/1983$) are: 272 $m^2/g$ total surface area, $a = 22\ Å$ calculated, $a = 28\ Å$ ($XRD$) and $a = 8 - 35\ Å$ ($TEM$). The particle sizes of both charges are not significantly changed after the reaction.

The difference $\Delta T$ between the catalyst bed temperature and the feedstock at the reactor inlet was measured by a $NiCr - Ni$ thermocouple, digitized and continuously recorded. The sampling rate is 2 points per second. 80 time series of that difference have been recorded scanning the values of the feed concentrations and temperature of the feedstock at the two charges of the catalyst.

At all 80 time series the difference $\Delta T$ exhibits irregular variations whose amplitude does not exceed 10% of the average "shift" of the catalyst bed temperature from the feedstock one. However, there are occasional large variations whose amplitude is about 50% of the average "shift".

In order to find out whether the features of these variations match the characteristics of a macroscopic evolution described by the type of eqs.(20) we have worked out the power spectrum of each recorded time series. Our study has shown that:

(i) all 80 power spectra comprise a continuous band of the shape $1/f^{\alpha(f)}$, where $\alpha(f) \to 1$ at $f \to 1/T$ ($T$ is the length of the sequence) and $\alpha(f)$ monotonically increases to $p > 2$ at $f \to \infty$. The shape is robust to the catalyst charge, feedstock temperature and the feed concentrations.

(ii) at certain values of the control parameters the power spectrum comprises a continuous band of the shape $1/f^{\alpha(f)}$ and a discrete band. Such coexistence is one of the hallmarks of a macroscopic evolution described by eqs.(20). Indeed, it justifies the additivity of the power spectrum - the discrete band comes from a system of ordinary differential equations (eqs.(21)) whereas the continuous band comes from the bounded stochasticity involved in eqs.(20). It should be stressed that this conclusion does not require any information about the particularities of the reaction mechanism and of the surface properties.

(iii) A large variation is developed at one of the time series. Its amplitude is about 50% of the average shift of the catalyst bed temperature from the feedstock one. The predominant amplitude of the variations is about 7%. In order to prove that the large variation induces a bifurcation we cut the time series into 3 successive parts: the first one involves the large variation, the second and the third one involve only "average" variations. The 3 parts are shown in Fig. 1a, 2a and 3a correspondingly. Their power spectra are presented in Fig.1b, 2b and 3b.

It is obvious that the presence of the large variation strongly affects the amplitude of the discrete band in Fig1a - it is about 10 times smaller than that of the discrete band in Fig.2b and 3b. On the other hand, the period remains the same at all 3 power spectra. The great sensitivity of the amplitude of oscillations to the distance from the bifurcation point along with robustness of the period to that distance is the typical property of a limit cycle. Thus, the large variation in Fig.1a causes effectively a strong "shift" of the control parameters though physically they have been sustained fixed. The impact of the large variation is temporary - it lasts as long as the variation is essentially large. Indeed, the amplitude of the limit cycle at Fig.2b and 3b is significantly different from that of the Fig. 1b.

The above results support the considerations of our previous subsection about the properties of the induced bifurcations - at each value of the control parameters an induced bifurcation needs a development of a fluctuation of certain size. The corresponding power spectrum always comprises a continuous band of the shape $1/f^{\alpha(f)}$. An induced bifurcation has a temporary effect - it lasts until the fluctuation is significant. The development of the large fluctuations is almost insensitive to the details of the reaction mechanism. That is why here we do not discuss the particularities of the reaction mechanism and the catalyst properties. They can answer questions such as: which is the particular bifurcation diagram, which are the thresholds of stability, does the variance of the fluctuations depend on the control parameters etc.

**Conclusions**

The major goal of the present paper is to elucidate the impact of the global and local boundedness on the structure and the properties of the state space of the natural systems. The boundedness is a necessary condition for keeping the evolution of the natural and artificial systems stable, namely: a long-term stability is possible if and only if the amount and the rate of energy and/or matter currently exchanged in any transition is bounded by the thresholds of stability of the system.

The relationship between the local and global boundedness and the stability of the system introduces two new general properties of the state space and the motion in it, namely: the state space is always bounded, the successive steps of motion are always finite and involve only nearest neighbors.

An immediate consequence of the global boundedness is that the invariant measure of the state space is the normal distribution. Thus, the global boundedness makes the normal distribution ubiquitous for the natural systems.

The existence of more than one accessible state to any given state makes each transition rate a multi-valued function. Indeed, at any instant only one transition takes place. However, it is arbitrarily chosen among the transitions to all accessible states. This introduces certain stochasticity of the transition rate even when all its selections can be carried out explicitly. The outcome is that the trajectory appears as a kind of a fractal Brownian "walk". The stochastisity induced by the choice of one selection among all available breaks any possible long-range periodicity (i.e. the incremental memory has finite size). So, the trajectories in our state space are fractal Brownian walks subjects to an incremental boundedness, namely: finite size of the increments (i.e. transition rates), finite size of the incremental memory and arbitrary selection rules.

In turn, the boundedness of the incremental memory size distinguishes a scale above which all the scales up to the thresholds of stability contribute uniformly to the creation of the fluctuations. Then, the coarse-grained structure of the attractor exhibits universal properties insensitive to the details of the incremental statistics. It has been established that the properties of each coarse-grained BIS, respectively a trajectory, are set on 3 specific parameters: the threshold of stability, the variance of the sequence and the exponent of the relation amplitude↔duration of a fluctuation. The non-zero but finite value of the latter provides finite velocity of the motion on any state space trajectory.

The condition that guarantees the asymptotic stability of the invariant measure (i.e. the normal distribution) reveals the interrelation between the boundedness as a necessary condition for a BIS (trajectory) to demonstrate chaotic properties and the presence of stretching and folding for keeping the evolution of a trajectory confined in a finite phase volume arbitrarily long time. From this viewpoint, an unlimited in the time stable evolution of a trajectory is possible if and only if the folding does not "feel" the boundary conditions imposed by the presence of the thresholds of stability. Otherwise, the evolution strongly depends on the way the excursion approaches the threshold.

We have proved that when certain relation (eq.(18)) among the major characteristics of a BIS holds, a folding with the desired properties exists. As a result, the presence of the thresholds is unperceptible that ensures all scales, thresholds included, uniformly to participate in setting the chaoticity of a BIS. In turn, the uniform participation of all scales, thresholds included, ensures an unlimited confinement of the BIS evolution in a finite attractor.

Consequently, on meeting eq.(18) a bounded state space subject to incremental boundedness appear as a global attractor whose invariant measure is the normal distribution. The universality along with the stationarity makes the attractor a transitive dense set of periodic orbits regardless to the particularities of the system. Thus, the structure of the attractor has steady properties insensitive to the development of any specific trajectory and/or the way of approaching the boundaries. The same properties render the motion on the attractor to exhibit the chaotic properties listed in the Introduction and established in (Koleva and Covachev, 2001).

Next a mechanism that gives rise to both local and global boundedness at each point of the state space is presented. The problem is particularly acute for the infinite systems where the lack of correlations among the local spatio-temporal fluctuations renders a self-sustained boundedness rather occasional and specific to a system or to certain states.

Though the presented mechanism gives rise to the chaotic kinetics introduced above, some issues remain open. Perhaps, one of the most important among them is the question about the modification of the synchronization in the presence of more than one type adspecies. This puts forward the issue about the formation and the stability of the spatio-temporal patterns as a process that starts at the synchronization.

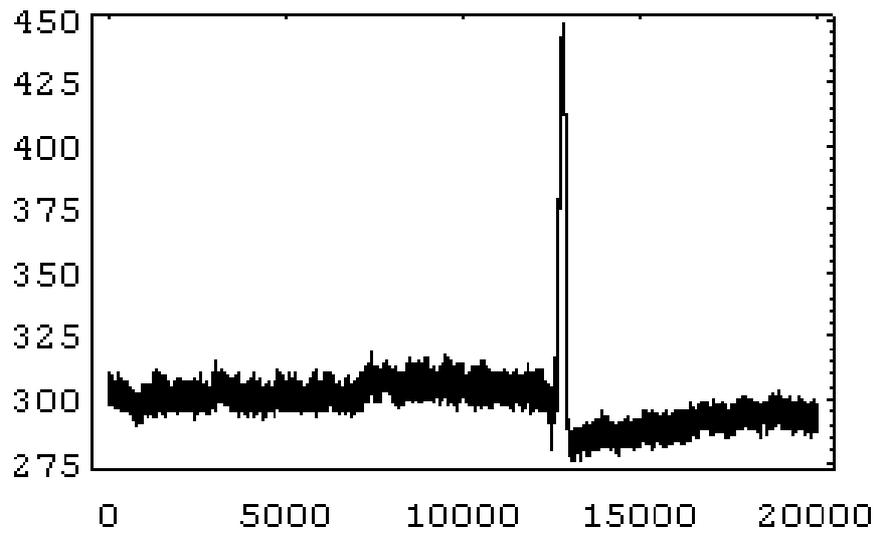

Figure 1a

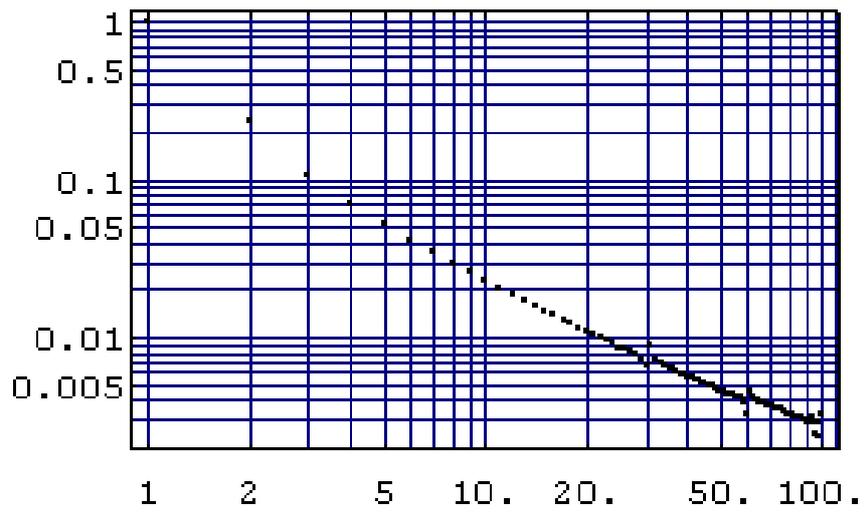

Figure 1b

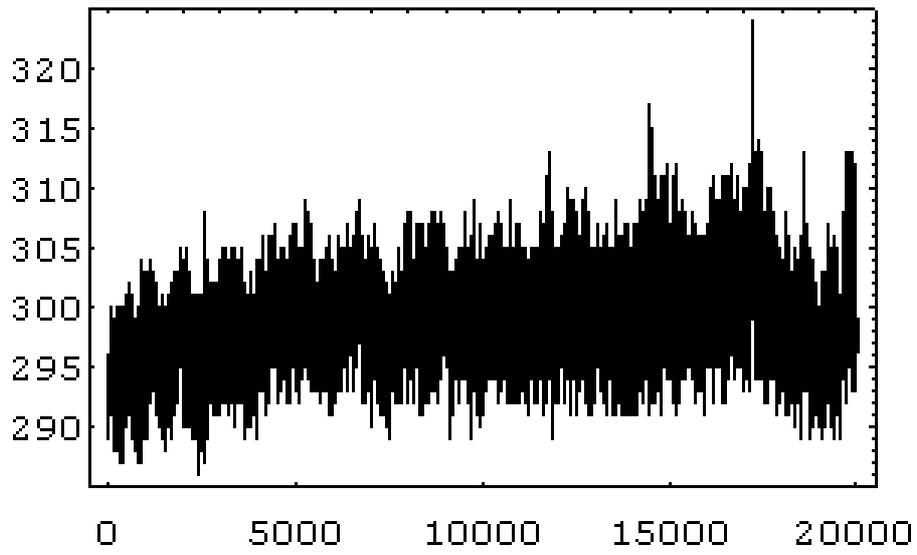

Figure 2a

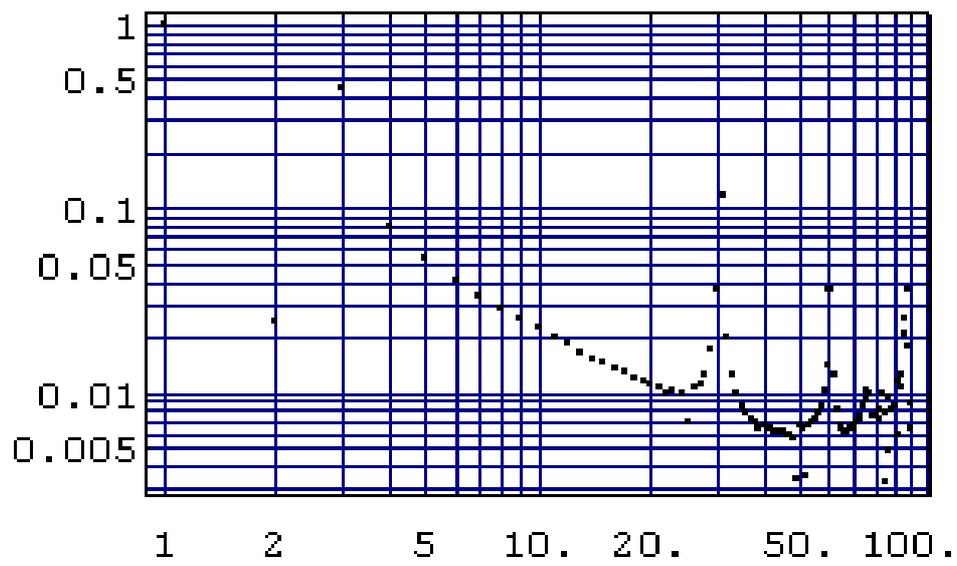

Figure 2b

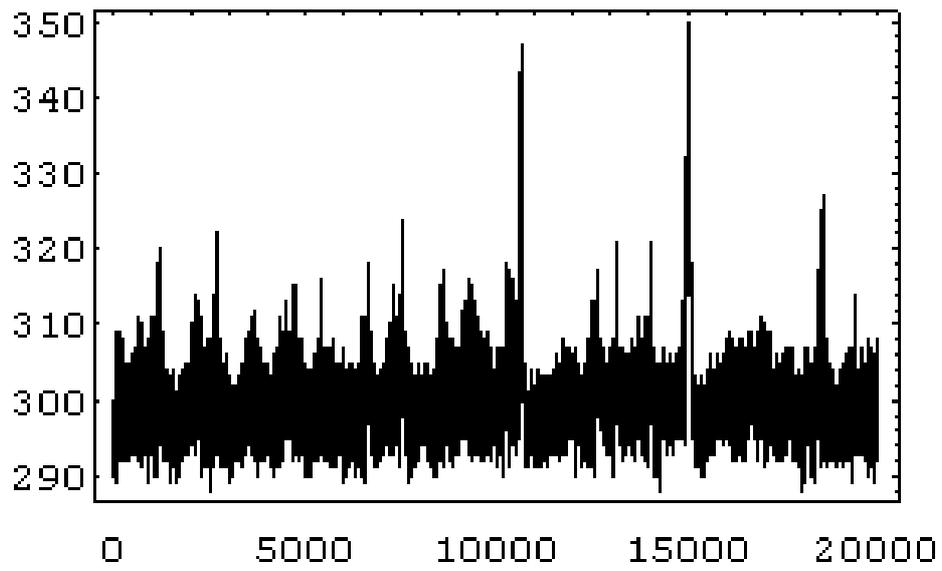

Figure 3a

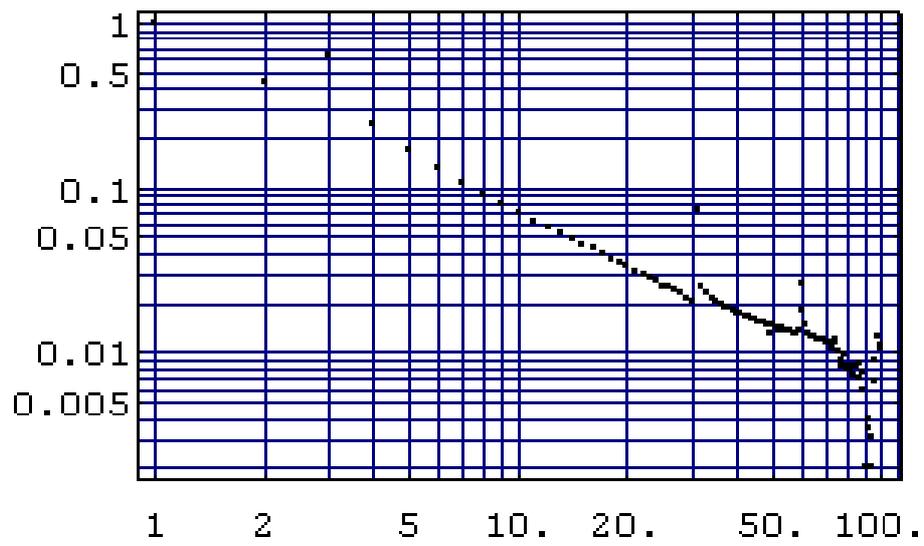

Figure 3b

# Captures to the Figures

Fig.1(a) A time series in relative units of the catalyst temperature variation (°K) in the course of time (sec) at the oxidation of HCOOH at a limit cycle: first part

Fig.1(b) A power spectrum in relative units of the catalyst temperature variations (°K) in the course of time (sec) at the oxidation of HCOOH at a limit cycle : first part

Fig.2(a) A time series in relative units of the catalyst temperature variation (°K) in the course of time (sec) at the oxidation of HCOOH at a limit cycle: second part

Fig.2(b) A power spectrum in relative units of the catalyst temperature variations (°K) in the course of time (sec) at the oxidation of HCOOH at a limit cycle : second part

Fig.3(a) A time series in relative units of the catalyst temperature variation (°K) in the course of time (sec) at the oxidation of HCOOH at a limit cycle: third part

Fig.3(b) A power spectrum in relative units of the catalyst temperature variations (°K) in the course of time (sec) at the oxidation of HCOOH at a limit cycle : third part